\documentstyle[preprint,aps]{revtex}
\input epsf.tex
\def\DESepsf(#1 width #2){\epsfxsize=#2 \epsfbox{#1}}
\begin{document}
\preprint{\vbox{\hbox{}}}
\draft
\title{
Signatures of Non-commutative QED at Photon Colliders}
\author{Seungwon Baek, Dilip Kumar Ghosh, Xiao-Gang He and W-Y. P. Hwang}
\address{
Department of Physics, National Taiwan University, 
Taipei, Taiwan.
}

\tightenlines

\date{ March, 2001}
\maketitle
\begin{abstract}
In this paper we study non-commutative (NC) QED signatures at
photon colliders 
through pair production of charged leptons$(\ell^+ \ell^-)$ and charged 
scalars$(H^+ H^-)$.
The NC corrections for the fermion pair production can be easily 
obtained since NC QED with fermions has been extensively studied in 
the literature. 
NC QED with scalars is less studied. To obtain the cross section for
$H^+H^-$ productions, we first investigate the structure 
of NC QED with scalars,
and then study the corrections due to the NC geometry to the ordinary
QED cross sections. Finally by folding in the
photon spectra for a $\gamma \gamma$ collider with laser back-scattered 
photons from the $e^+ e^-$ machine, we obtain $95\%$ CL lower 
bound on the NC scale using the above two processes. 
We find that, with $\sqrt{s}
= 0.5,\;1.0$, and $\;1.5$~TeV and integrated luminosity $L = 500(fb^{-1})$, 
the NC scale up to 0.7, 1.2, and 1.6 TeV can be probed, respectively,
while, for monochromatic photon beams, these numbers become 1.1, 1.7, 
2.6 TeV, respectively.
\end{abstract}


\preprint{\vbox{\hbox{}}}


\section{Introduction}

The property of space-time has fundamental importance in
understanding the law of nature. Non-commutative (NC) quantum
field theory provides an alternative to the ordinary quantum field
theory which may shed some light on the detailed structure of
space-time and have been studied in the past~\cite{snyder}. 
Recently NC quantum field theory and its applications
has also developed within string theories where it arises 
in low energy excitations of D-branes in the presence of certain
$U(1)$ background field and has received a lot of
attention~\cite{nc_string}. A simple way to modify the commutation relation
for the ordinary space-time $x$ is defined, with the modified 
space-time coordinate $\hat X$, as

\begin{eqnarray}
[\hat X_\mu, \hat X_\nu] = i \theta_{\mu\nu} = {i\over \Lambda^2}
c_{\mu\nu}.
\label{eq:NCgeo}
\end{eqnarray}
In the above the parameter $\Lambda$ which has the dimension of energy
signifies the scale where NC effects become relevant.
$c_{\mu\nu}$ is a real anti-symmetric matrix with elements of order
one which commute with the space-time coordinate $x^\mu$. 

Phenomenologically 
the NC scale $\Lambda $ can take any value, the likely one being
of the order of the Planck scale $\overline{\rm M_{Pl}}$. However,
the recent studies in the area of large extra dimensions show
that gravity becomes strong at the TeV scale~\cite{extra}, and also one might
see some stringy effects at this scale. Hence, it is 
justified if one takes the scale of $\Lambda$ to be of the order
of TeV scale. If this is the case, then whether NC geometry has 
anything to do with reality has to be tested experimentally. 
In this context the Next Generation $e^+e^-$ Linear Collider(NLC) 
will be an ideal machine to probe such new physics effects. The 
$e^+e^-$ version of NLC can be modified to give $e^-e^-$, $e\gamma$
and $\gamma\gamma$ mode of collider. Some of the authors have already 
studied the NC effects at NLC~\cite{arfaei,rizzo}. 

In this paper we study signatures of NC Quantum Electrodynamics (QED) 
at $\gamma \gamma$ colliders. $\gamma\gamma $ colliders can be very 
sensitive to certain new physics beyond the standard model~\cite{jika}.
We also find that $\gamma\gamma$ colliders can provide interesting information
about the scale $\Lambda$ of non-commutative geometry. Two processes
$\gamma\gamma\to \ell^+\ell^-$ and $\gamma\gamma \to H^+ H^-$ will be studied 
in detail. These processes are particularly interesting in studying the
non-commutative QED effects because at leading order they are
purely QED processes, eliminating problems associated with difficulties
to have a full gauge theory for $SU(3)_C\times SU(2)_L \times
U(1)_Y$. This is because that only U(N) group can be gauged
consistently with NC geometry~\cite{matsubara}. 
The gauge group of the standard model has to be enlarged in the presence
of NC geometry, Eq.~(\ref{eq:NCgeo}).
If weak interaction is involved,
then there is problem to identify NC effects such as the process
$e^- e^- \to e^- e^-$ where exchange of Z boson also contributes~\cite{rizzo}.

The paper is organized as follows. In section II we study the 
$\gamma\gamma \to \ell^+\ell^-$ process in NC QED with monochromatic
photon beams and laser back-scattered photon beams for three values of
center-of-mass energies $0.5$~TeV, $1$~TeV and $1.5$~TeV~\cite{nlc_rev}. 
We obtain $95\%$ CL lower bound which can be probed on the NC scale 
$\Lambda $. 
In section III we study
the non-commutative scalar QED. We first derive the corresponding 
Feynman rules and use them to obtain $95\%$ CL lower bound on $\Lambda$ from
$\gamma\gamma \to H^+ H^-$ process. 
Finally in section IV, we summarize our results.

\section{ $\gamma\gamma \to \ell^+ \ell^-$ IN NC QED}
In this section we study the effects of NC QED in the 
$\gamma\gamma \to \ell^+ \ell^-$ process.
NC QED with fermions has been studied extensively~\cite{arfaei}.
The Feynman rules relevant are shown in Fig.~1 and the Feynman diagrams 
for $\gamma\gamma \to \ell^+ \ell^-$ are shown in Fig.~2. It is 
clear from the Feynman rules and as well as from Fig.~2 that there are 
extra contributions to the ordinary QED. The ordinary 
QED vertex is modified to have a momentum dependent phase factor. Apart from
this there are completely new triple and quartic photon 
vertices making the NC QED 
like a non-abelian gauge theory. The origin of phase factors in the 
vertices can be traced back to the famous Weyl-Moyal 
correspondence~\cite{riad} which we 
will state later. These new contributions to 
the existing vertices result in
deviations from the ordinary QED predictions.  We obtain the unpolarized 
differential cross section for 
$\gamma(k_1)\gamma(k_2) \to \ell^-(p_1) \ell^+(p_2)$ process 
in the massless limit, as
\begin{eqnarray}
{d\sigma\over dz d\phi} = {\alpha^2\over 2 \hat s}
\bigg[{\hat u\over \hat t} + 
{\hat t\over \hat u} - 4 {\hat t^2+\hat u^2\over s^2}
 \sin^2 \delta \bigg],
\end{eqnarray}
where the NC phase is $\delta = (\frac{k_1\cdot\theta\cdot k_2}{2})$. 
$\hat s=(k_1+k_2)^2=(p_1+p_2)^2$, $\hat t=(k_1-p_1)^2 = (k_2-p_2)^2$
and $\hat u = (k_1-p_2)^2 = (k_2-p_1)^2$ are the standard Mandelstam variables.
In the $\gamma\gamma $ center-of-mass frame,
$\hat t$ and $\hat u$ can be further written in terms of 
$\hat s$ and the angle $\hat{\theta}$ 
between $\vec k_1$ (the z-direction) and $\vec p_1$ with
 $\hat t = -\frac{\hat s}{2}(1-z)$, $\hat u=-\frac{\hat s}{2}(1+z)$, 
where $z = \cos \hat{\theta}$. The angle  $\phi$ is the azimuthal angle.
So, the NC effect in the $\gamma\gamma \to \ell^+\ell^-$ process lies in the
even function $\sin^2\delta$ of $\delta$ and one can recover the ordinary QED 
result by taking the 
limit $\delta \rightarrow 0 $. The phase $\delta$ arises from the $s$-channel 
triple photon vertex diagram and also from the interference between the 
$s t$-channel and $s u$-channel diagrams.

The cross sections are only sensitive to the
NC parameter $c_{0z}$ because the corrections only depend
on $\sin^2({1\over 2}k_1\cdot \theta \cdot k_2)$ which is equal to
$\sin^2[(\hat{s}/4)(c_{0z}/\Lambda^2)]$. 
Because of this the cross section does not depend on the azimuthal 
angle $\phi$ and will be integrated over in the cross section from now on.
In our later discussions, we will
set $c_{0z} =1$ and study the sensitivity to the NC scale $\Lambda$.

Now we study the cross-section as a function of the NC scale $\Lambda$.
In obtaining the cross-section we sum over three leptonic ($e,\mu,\tau$)
generations. We also assume that the identification efficiencies for $e$ 
and $\mu$, and $\tau$ to be $100\%$ and $60\%$ respectively. One should note
that, due to the neglect of small lepton masses, there are singularities in 
the cross section  when $z=\pm 1$. To avoid these singularities, we demand that 
the rapidity $\mid \eta_{\ell}\mid $ of each lepton should be less than 1. 
This choice of rapidity cut corresponds to an angular cut $\mid z\mid < 0.76 $ 
on each leptons.
In Fig. 3 we plot the variation of this cross-section with $\Lambda$ for
a monochromatic (line with $\odot$ ) 
photon collider with energy $\sqrt{s}_{\gamma\gamma}=1$ TeV. 
The adjacent solid line represents ordinary QED cross-section. 
It can be seen that the ordinary QED gets negative
contribution from NC QED, and as the NC scale $\Lambda$ increases the
NC QED result asymptotically approaches to the ordinary QED one.

To study the possible sensitivity of NLC to the NC scale $\Lambda$ we perform
$\chi^2(\Lambda)$ fit assuming that statistical errors are Gaussian and that
there are no systematic errors. $\chi^2(\Lambda)$ is given by
\begin{eqnarray}
\chi^2 =  L {(\sigma_{NC}(\Lambda) - \sigma_{SM})^2\over \sigma_{SM}},
\end{eqnarray}
where $L$ is the integrated luminosity, $\sigma_{SM}$ is the 
ordinary QED total
cross-section and $\sigma_{NC}(\Lambda)$ is the NC QED cross-section.
By demanding $\chi^2 \geq 4$, we obtain the lower bound 
on the NC scale $\Lambda$ at $95\%$ CL. We denote this bound by 
$\Lambda^{\rm lower}$.
We take three machine energies $\sqrt{s}_{e^+e^-} = 0.5$, 
1.0 and 1.5~TeV, for illustrations. In the case of monochromatic photon
collider, $\sqrt{s}_{\gamma\gamma} = \sqrt{s}_{e^+e^-}$.

In Fig. 4 we show the scale $\Lambda^{\rm lower}$ as a function
of $L$ from $\gamma\gamma \to \ell^+\ell^-$
process. The solid lines in Fig. 4 represent 
as a function of integrated 
luminosity $L$ for monochromatic photon beams. Higher the 
$\sqrt{s}_{e^+e^-}$ larger the value of $\Lambda$ can be probed for a 
fixed integrated luminosity. This behavior can be understood from the nature of
the NC correction term which goes as $\sim s/\Lambda^2$ to this process.  

Till now, we have discussed about the monochromatic photon beams. However, it
is very difficult to obtain such a beam in practice.
A realistic method to obtain high energy photon beam is 
to use the laser back-scattering technique on an electron 
or positron beam which produces abundant 
hard photons nearly along the same direction as the original
electron or positron beam. The photon beam energy obtained this way is
not monochromatic. The energy spectrum of the back-scattered photon is
given by~\cite{ginzburg}
\begin{eqnarray}
f(x)&=& {1\over D(\xi)} [1-x+{1\over 1-x} - {4x\over \xi(1-\xi)} +
{4x^2\over \xi^2(1-x)^2}],\nonumber\\
D(\xi)&=& (1-{4\over \xi}-{8\over \xi^2})\ln(1+\xi) + {1\over 2} +
{8\over \xi} - {1\over 2(1+\xi)^2}.
\end{eqnarray}
where $x$ is the fraction of the energy of the incident $e^\pm$ beam. The
parameter $\xi$ is determined to be $2(1+\sqrt{2})$ by requiring that 
the back-scattered photon to have the largest possible energy, but does not 
interfere with the incident photon to create unwanted $e^+e^-$ pair which
sets $x_{max} = \xi/(1+\xi) \approx 0.828$.
The cross section at such a $\gamma \gamma$ collider with the 
$e^+ e^-$ collider center of mass frame energy
$\sqrt{s}$ is given by

\begin{eqnarray}
\sigma = \int^{x_{max}}_{x_{1min}}dx_1 f(x_1)
\int^{x_{max}}_{x_{2min}} dx_2 f(x_2)\int^{z_{max}}_{z_{min}} dz
{d\sigma(x_1x_2 s, z)\over dz}.
\end{eqnarray}

To avoid the singularities at $z=\pm 1$ in $\gamma \gamma\to \ell^+ \ell^-$, 
we make a cut on the rapidity of each lepton 
in the laboratory frame to be less than 1 and also a cut on the lepton
energy such that
the minimal values for $x_{1,2}$ to be
$x_{1,2min} =0.5$.
With this choice of cuts we show the variation of cross-section 
with $\Lambda$ in Fig. 3. The dotted lines represent the NC QED cross-section,
while the adjacent solid lines correspond to the ordinary QED results.  
Compared to the monochromatic case, the cross-section decreases. Naively one
would expect the other way around
because the cross-section decreases with energy. 
However, due to the cut on $x_{1,2min}$, certain portion of the scattering 
is also cut off which results in a smaller cross-section. 

In Fig. 4  we present $\Lambda^{\rm lower}$
as a function of $L$ with dashed line. In this case, for a given 
$\sqrt{s}$ and integrated luminosity, the $95\%$ CL lower bounds are
weaker than that of monochromatic photons.
For example, at $\sqrt{s}_{e^+e^-} =1$ TeV 
and assuming the integrated luminosity  
$L = 500~{\rm fb}^{-1}$ the $95\%$ CL lower bound can be probed 
on $\Lambda$ is 1.2 TeV.
While in the monochromatic case, the corresponding bound is 1.6 TeV.
This is due to the fact that the  available $\gamma\gamma $ 
center-of-mass energy is not fixed but has an energy spectrum,
which suppresses the NC effect.

\section {$ \gamma\gamma \to H^+ H^- $ in NC scalar QED}

NC QED with scalars are less studied. In order to study $\gamma
\gamma \to H^+ H^-$, we first construct the NC QED Lagrangian with
scalars in the following. The Lagrangian in the ordinary quantum
field theory relevant to $\gamma\gamma \to H^+H^-$ is given by

\begin{eqnarray}
L = (D^\mu H^-)^* (D_\mu H^-) - m^2_H H^+ H^-,
\end{eqnarray}
where $D_\mu = \partial_\mu - i e A_\mu$ is the covariant
derivative.

When the above Lagrangian is formulated with non-commutative coordinates,
there are corrections.
NC quantum field theory can be easily studied using the Weyl-Moyal
correspondence replacing the product of two fields
$A(\hat X)$ and $B(\hat X)$ with NC coordinates by~\cite{riad}

\begin{eqnarray}
A(\hat X) B(\hat X) \to A(x)*B(x) = [e^{{i\over 2}
\theta_{\mu\nu}\partial_{x}^\mu\partial_{y}^\nu} A(x)
B(y)]_{x=y},
\end{eqnarray}
where $x$ and $y$ are the ordinary coordinates, and 
$\partial_x = \partial/\partial x$, $\partial_y = \partial/\partial y$.

Under an infinitesimal local gauge transformation $\lambda(x)$, the
transformation law for $H$ is given by

\begin{eqnarray}
ReH^-(x) &\to& ReH^-(x) - \cos ({1\over 2} \theta_{\mu\nu}
\partial_{x}^{\mu} \partial_{y}^{\nu})
 \lambda(x) Im H^-(y)|_{x=y}\nonumber\\
&-&\sin({1\over 2}\theta_{\mu\nu}
\partial_{x}^{\mu}\partial_{y}^{\nu}) \lambda(x) Re
H^-(y)|_{x=y},\nonumber\\ 
Im H^-(x) &\to& Im H^-(x) +\cos({1\over 2}
\theta_{\mu\nu}\partial_{x}^{\mu}\partial_{y}^{\nu}) \lambda(x) Im
H^-(y)|_{x=y}\nonumber\\
& -& \sin({1\over 2}
\theta_{\mu\nu}\partial_{x}^{\mu}\partial_{y}^{\nu}) \lambda(x)
ReH^-(y)|_{x=y}.
\end{eqnarray}

Writing the Lagrangian in NC geometry, one obtains the tree level
NC QED with scalars. We have

\begin{eqnarray}
L = (\partial_\mu H^+ + ie H^+*A_\mu)*(\partial^\mu H^- - ie A^\mu * H^-)
- m^2_H H^+*H^-.
\end{eqnarray}
Due to the NC properties, the ordering of the fields in the above
equation is important and should not be misplaced.
From this Lagrangian one obtains the Feynman rules given in Fig. 5. The 
structure shows some similar momentum dependent phase factor as in the
NC QED with fermions.  
The Feynman diagrams for $\gamma\gamma\to H^+ H^-$ are shown in Fig. 6. 
In this case also we get an additional contribution to the normal QED
process from an extra $s$-channel diagram, which has a non-abelian kind
of structure. The amplitude for this process can be expressed as
\begin{eqnarray}
iM &=& 2 i e^2 \epsilon^\mu(k_1) \epsilon^\nu(k_2) e^{{i\over 2}p_1\cdot
\theta\cdot p_2}\nonumber\\
&\times& \bigg[
{ i\over s} \sin\left({k_1\cdot \theta\cdot k_2 \over 2}\right)
\Big((u-t)g_{\mu\nu} + 
2 k_{2\mu}
(p_1-p_2)_\nu - 
2 k_{1\nu}
(p_1-p_2)_\mu \Big)\nonumber\\ 
&+& 
\cos\left({k_1\cdot \theta\cdot k_2 \over 2}\right) g_{\mu\nu} +
e^{-{i\over 2} k_1\cdot \theta\cdot k_2}
{2 p_{1\mu} p_{2\nu}\over t -m^2_H} 
+ e^{{i\over 2}k_1\cdot\theta\cdot k_2}
{2 p_{2\mu} p_{1\nu}  \over u-m^2_H} 
\bigg].
\end{eqnarray}

In obtaining the cross section one should be careful about
the non-abelian nature of the triple photon vertex since more than one
gauge bosons are involved, that is one should treat the photon polarization
sum with care to make sure that Ward identities are satisfied and also 
to guarantee that the unphysical photon polarization states do not appear. 
We have worked with two methods with the same final results, 
one using explicit
transverse photon polarization vectors, and another using~\cite{book} 
\begin{eqnarray}
\sum_\lambda \epsilon^\mu(\lambda) \epsilon^{\nu*}(\lambda)
=-\bigg [g^{\mu\nu} -\frac{n^\mu k^\nu+n^\nu k^\mu}{(n.k)}
+\frac{n^2k^\mu k^\nu}{(n.k)^2} \bigg ].
\end{eqnarray}
where $n$ is any arbitrary $4$-vector and $k$ is the photon $4$-momentum. 
The same technique has been applied in the above charged lepton pair
production study.
The unpolarized differential cross section in the 
$\gamma\gamma $ center-of-mass frame is given by 
\begin{eqnarray}
   {d \sigma \over dz d\phi} &=& {\alpha^2 \hat{\beta} \over 4 \hat{s}} \;
    \bigg[
     \frac{(m_H^2 + \hat{t})^2}{(m_H^2 - \hat{t})^2}	
    +\frac{(m_H^2 + \hat{u})^2}{(m_H^2 - \hat{u})^2}
    +\frac{8m_H^4}{(m_H^2 - \hat{t})(m_H^2 - \hat{u}) } \bigg] \nonumber\\
   &\times&\bigg[ 1-4 \frac{(m_H^2 - \hat{t})(m_H^2 - \hat{u})}{\hat{s}^2}
     \sin^2\delta \bigg].
\end{eqnarray}
where the NC phase $\delta$ has been defined earlier,
$\hat t = m^2_H - \frac{\hat s}{2}(1 - \hat \beta z) $,
$\hat u = m^2_H - \frac{\hat s}{2}(1 + \hat \beta z) $, and
$\hat \beta = \sqrt{1 -4 m^2_H/\hat s}$ is velocity of the charged scalar.
In this case also, in the limit $\delta \rightarrow 0$, one obtains the
pure QED result. Again this process depends only on $c_{0z}/\Lambda$.

The scalars are similar to charged Higgses in multi-Higgs models.
However the decay products are not clear because the minimal Standard Model
for electroweak interactions have to be extended with NC geometry. The
charged scalar decay products may be modified. We will assume that the decay
products of $H$ are similar to the charged Higgs scalars in multi-Higgs models 
and can be studied experimentally.
One may also formulate NC QED with composite charged scalars, such as
$\pi^\pm$ and $K^\pm$ which will be commented on later. 

The variation of unpolarised cross-section with $\Lambda$ for
scalar mass $m_H = 100$~GeV at $\sqrt{s}_{\gamma\gamma}=1$ 
TeV is also shown in
Fig. 3 for both monochromatic (line with dark boxes) 
and laser back-scattered (dashed lines) photon beams. The corresponding 
ordinary QED contributions, are also depicted by the solid lines. 
From this figure it can be seen that the ordinary QED gets negative
contribution from NC QED like the $\gamma \gamma \to \ell^+ \ell^-$ process
, and as the NC scale $\Lambda$ increases the NC QED contribution 
asymptotically approaches to the ordinary QED result. 
It is interesting to note that the cross-section in the back-scattered case is
larger than that of monochromatic one, unlike the $\gamma\gamma\to \ell^+ \ell^-$
case discussed earlier. This is because that in this case no cut on 
the final product energy is applied, therefore all contributions are included.
However, the monochromatic case still has larger 
deviation between the ordinary and non-commutative QED as can be seen 
in Fig. 3.

Now we discuss our results on $\Lambda^{\rm lower}$
for $\gamma \gamma \to H^+ H^-$ with monochromatic photon beams. 
We use two charged scalar masses,
$m_H = 100$~GeV and $200$~GeV, for illustrations.
We display $\Lambda^{\rm lower}$
as a function of $ L$ by the solid lines in the Fig.7 (a)
($m_H= 100$~GeV) and  Fig.7 (b)($m_H = 200$~GeV). The numbers adjacent to each 
curve correspond to monochromatic photon collider.
It is clear from these
two figures that 
$\Lambda^{\rm lower}$ does 
depend on the scalar mass. The lighter the mass,
the larger the scale one can explore
for a given $\sqrt{s}$ and integrated luminosity. 
For example, with $\sqrt{s}_{\gamma \gamma}=1$ TeV, 
$\Lambda^{\rm lower}$ are 1.53 TeV for $m_H = 100$~GeV 
and $1.48$~TeV for $m_H = 200$~GeV, respectively.
Like the dilepton final state, here also, one can probe larger 
value of $\Lambda$ if one goes to higher energies.

The results for the laser back-scattered photon beams are also shown 
in Figs. 7(a) and 7(b) by the dotted lines. 
In the case, there are no singularities at $z=\pm 1 $. Therefore we will 
let $z$ vary within the full allowed range,
that is with $z_{min} = -1, z_{max} = 1 $. The integration lower limits for 
$x_1$ and $x_2$ are: $x_{1min} = 4m^2_H/sx_{max}$
and $x_{2min} = 4m^2_H/sx_1$. The maximum value of $x_1$ and $x_2$ has been
already mentioned in section II. Using Eqn.(3) we then obtain 
$\Lambda^{\rm lower}$ as a function of integrated luminosity $L$ which 
is shown by the dotted lines in Fig. 7. As before, we study this case also 
for two values of scalar masses 100 GeV and 200 GeV. 
We see that the bounds which can be probed on the scale 
are in the range of  0.8 to 1.2, 0.7 to 1.0, and 0.4 to 0.6 TeV for
$\sqrt{s}_{\gamma\gamma} =1.5$, 1.0, and 0.5 TeV, respectively. 
These
bounds are slightly lower than that 
obtained in $\gamma\gamma\to
\ell^+\ell^-$.
 
If the theory  is applicable to
composite particles such as $\pi^\pm$ and $K^\pm$, 
the NC scale that can be probed with $L=500$ fb$^{-1}$ is
1.5 TeV (1.2 TeV) for monochromatic (back-scattered) photon beams for
$\sqrt{s}_{\gamma\gamma}=1$ TeV ($\sqrt{s}_{e^+ e^-}=1$ TeV).
Of course it may be difficult to carry out such experiments with
energies as high as what we are considering. 

In the above discussions we have used tree level cross sections, especially
our reference ordinary cross sections $\sigma_{SM}$. There are loop
contributions which may lead to the change of $\chi^2$ compared with
when tree cross sections
are used. However, the loop corrections are much smaller than the
NC corrections for $\chi^2$ as large as 4. The bounds we obtained are
for NC corrections to good approximations.

\section{Conclusions}
In summary, we have examined the feasibility of observing the 
experimental signature of non-commutative QED by studying 
dilepton and pair of charged scalars productions at high energy
photon collider. We have parametrized the effect of NC QED by 
an anti-symmetric matrix  $c_{\mu\nu}$ and an overall NC scale 
$\Lambda $. We found that in our processes only $c_{0z}$ contributes.
Throughout our analysis we have set this parameter to 1 and studied the
sensitivities of $\gamma\gamma \to \ell^+\ell^-$ and
$\gamma\gamma \to H^+ H^-$ processes on the NC scale $\Lambda$.

We first studied the sensitivity for
monochromatic $\gamma\gamma $ colliders. 
The variation of $\sigma (\gamma\gamma \to \ell^+ \ell^-)$
and $\sigma (\gamma\gamma \to H^+ H^-)$ with the NC scale $\Lambda$ at
$\sqrt{s}_{\gamma\gamma}= 1$ TeV were obtained. 
We found that there are visible deviations between ordinary and NC QED
predictions for small $\Lambda$, but when $\Lambda$ becomes larger, 
$\sigma_{NC}$ asymptotically approaches $\sigma_{SM}$.
We also obtained $95 \%$ CL lower limit which can be probed
on $\Lambda$ from above mentioned
two processes as functions of the integrated luminosity $L$. It turned out 
that higher the available center-of-mass energy larger the
NC scale $\Lambda$ one can probe.
We found that with $\sqrt{s}_{\gamma\gamma} = 0.5,\;1.0$, and $\;1.5$~TeV and 
integrated luminosity $L = 500(fb^{-1})$
the NC scales can be probed up to 1.1, 1.7, and 2.6 TeV,
respectively. 
 
Next we considered more realistic case, where, the photon beams
are obtained by laser back-scattered from $e^\pm$ beams.
In this case, the available $\gamma \gamma $ center-of-mass 
energy has an spectrum with a maximum energy around $80\%$ of the
$\sqrt{s}_{e^+e^-}$. In general bounds on the scale that can be probed
become lower. 
We have observed that for $ \sqrt{s}_{e^+e^-}
= 0.5,\;1.0$, and $\;1.5$~TeV, with the integrated luminosity 
$L = 500(fb^{-1})$, the NC scales up to 0.7, 1.2, and 1.6 TeV can be probed, 
respectively. 

In both monochromatic and laser back scattered photon collider cases, the bounds
on $\Lambda$ can be probed using $\gamma\gamma \to H^+ H^-$ are slightly
lower than that can be obtained using $\gamma\gamma\to \ell^+\ell^-$. 

\acknowledgements  
This work was supported in part by National Science Council 
under the grants NSC 89-2112-M-002-016 and NSC 89-2112-M-002-062, and
in part by the Ministry of Education Academic Excellent Project
89-N-FA01-1-4-3.
\newpage
\begin{figure}[htb]
\centerline{ \DESepsf(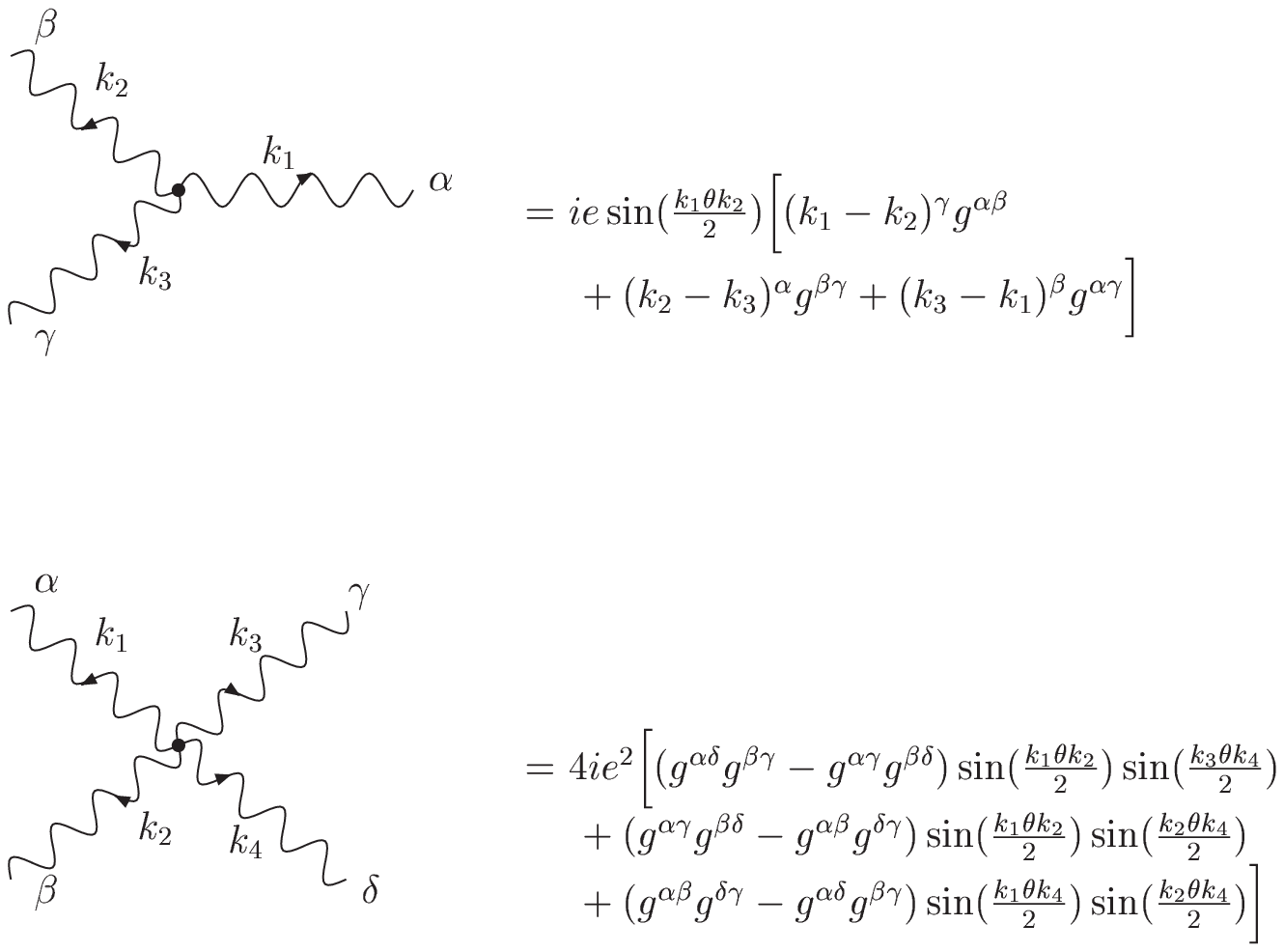 width 20cm)}
\vspace*{-6.in}
\caption { Feynman rules for NC QED with fermions.  }
\end{figure}
\newpage
\begin{figure}[htb]
\centerline{ \DESepsf(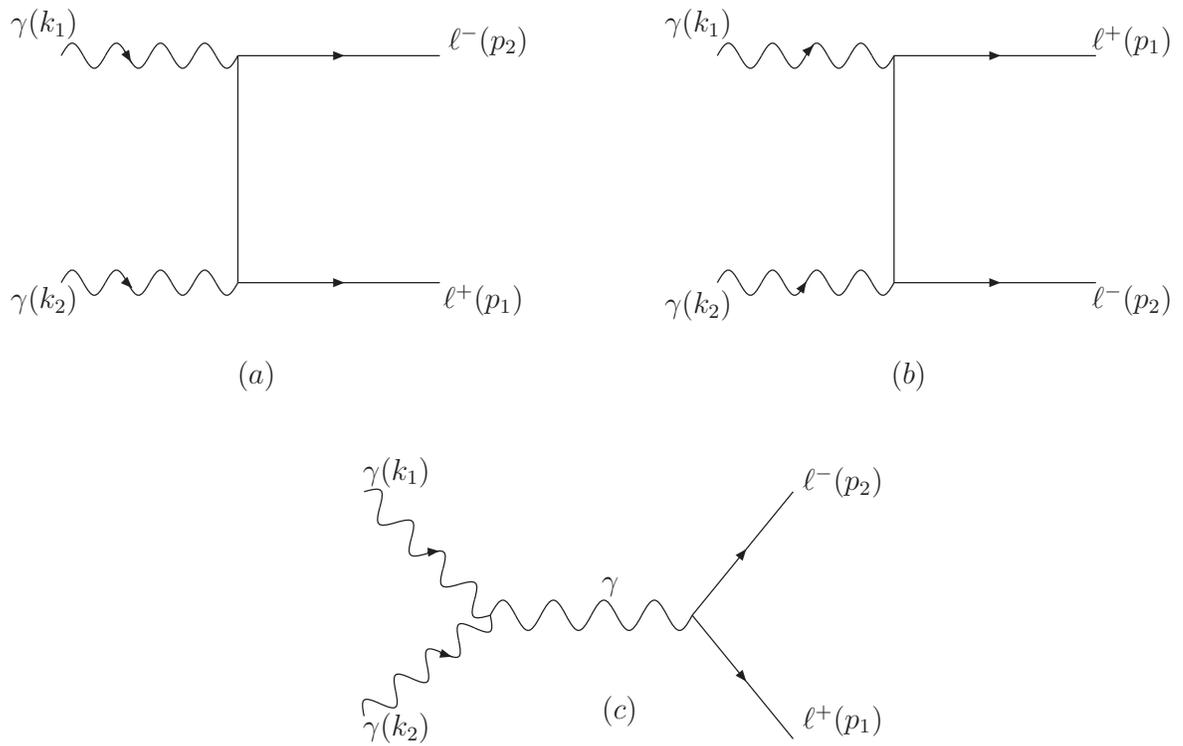 width 20cm)}
\vspace*{-6.in}
\caption { Feynman diagram for $\gamma\gamma \to \ell^+\ell^-$ in the
presence of NC QED.}
\end{figure}
\newpage
\begin{figure}[htb]
\centerline{ \DESepsf(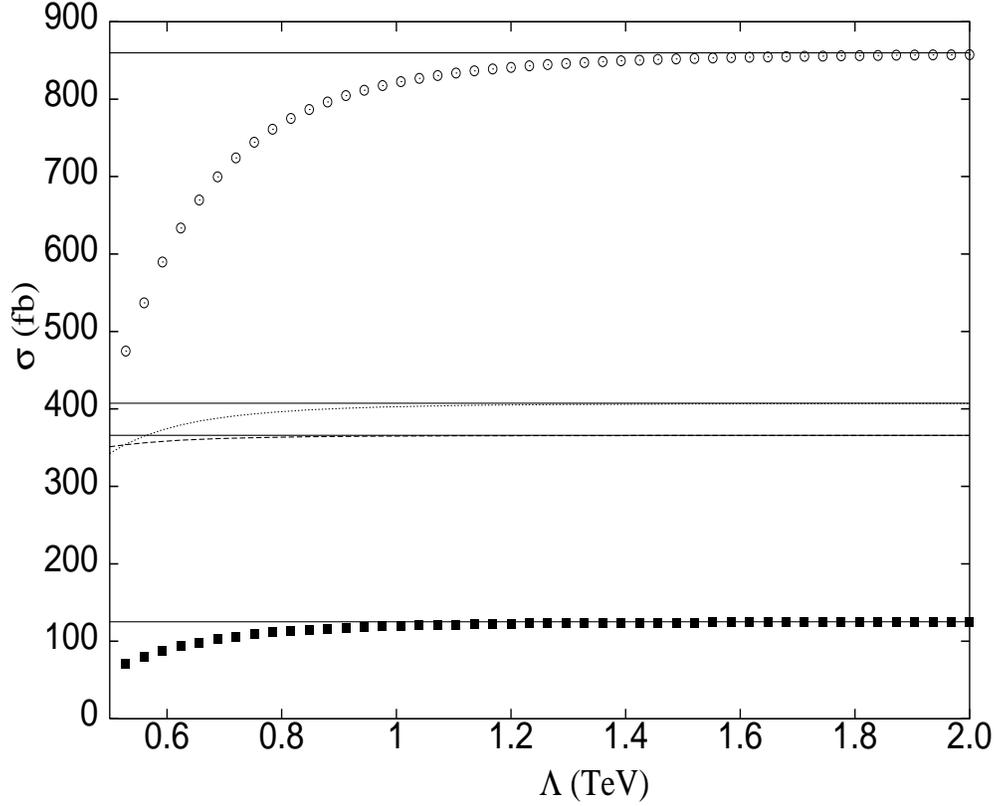 width 14cm)}
 \vspace*{-2.in}
\caption { Variation of $\sigma (\gamma\gamma \to \ell^+ \ell^-) $ and 
$\sigma (\gamma\gamma \to H^+ H^-) $ with the NC scale $\Lambda$ 
at $\sqrt{s}_{e^+e^-} =1$~ TeV NLC machine. The notations are following:
$(i)$ $ \sigma (\gamma\gamma \to \ell^+ \ell^-) $ with 
monochromatic and with laser back-scattered photon beams 
are represented by the curve with $ \odot $ and with dotted lines respectively. 
The solid lines adjacent to these correspond to commutative QED 
contribution. $(ii)$ $\sigma (\gamma\gamma \to H^+ H^-) $ with
monochromatic and with laser back scattered photon beams 
are represented by the curve with dark boxes and with dashed lines 
respectively. The solid lines adjacent to these correspond to ordinary QED 
contribution. For this we have fixed the scalar mass $m_H = 100$~GeV.}  
\end{figure}

\newpage
\begin{figure}[htb]
\centerline{ \DESepsf(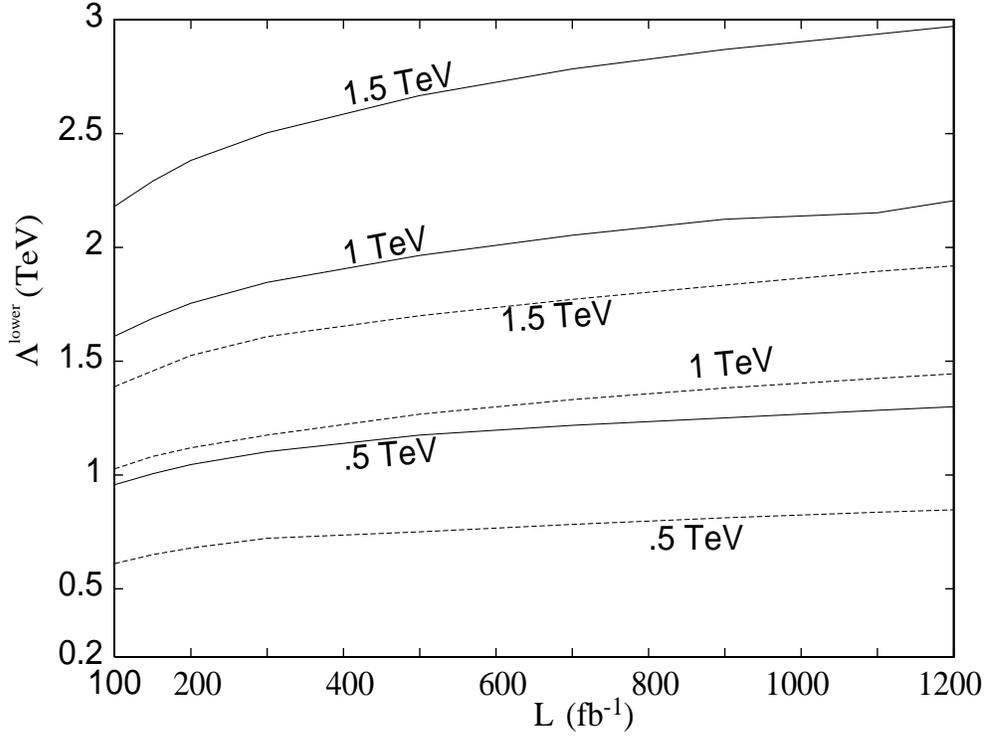 width 12cm)}
\smallskip
\caption {$95\%$ CL lower bound on $\Lambda $ can be probed 
as a function of integrated 
luminosity from $\gamma\gamma \rightarrow \ell^+ \ell^-$ process.  
The solid lines are using monochromatic photon beams, while the dashed 
lines are with back-scattered photons. The numbers adjacent to each curve 
represents the $\sqrt{s}_{e^+e^-}$. }
\end{figure}
 \begin{figure}[htb]
 \centerline{ \DESepsf(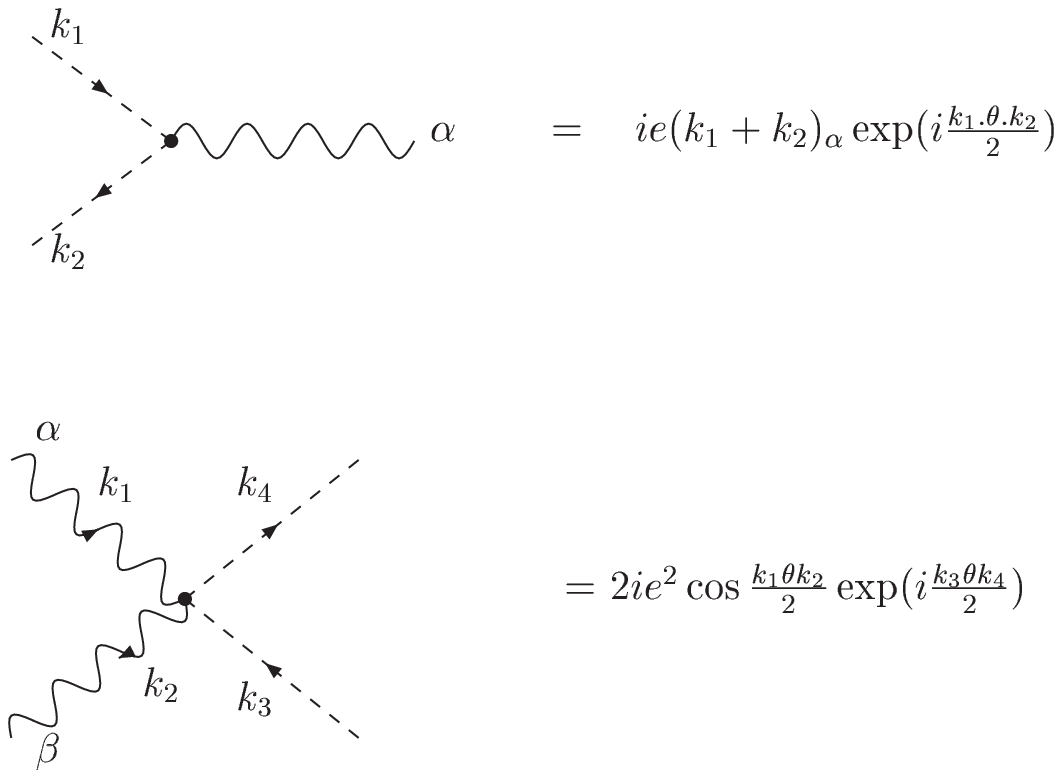 width 20cm)}
 \vspace*{-6.in}
 \caption {  Feynman rules for NC scalar QED.}
 \end{figure}
\newpage
\begin{figure}[htb]
\centerline{ \DESepsf(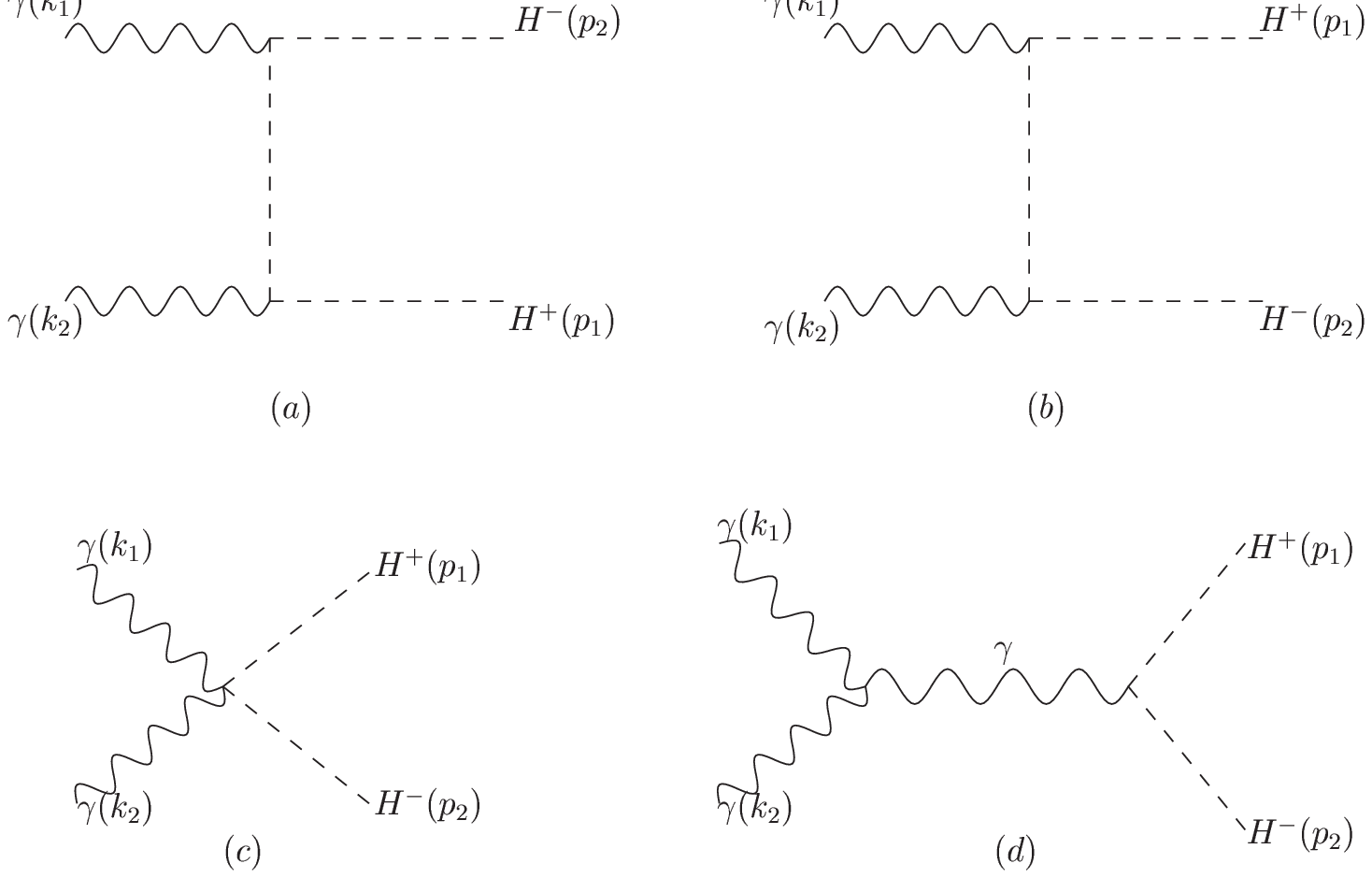 width 20cm)}
\vspace*{-6.in}
\caption { Feynman diagram for $\gamma\gamma \to H^+ H^-$ in the
presence of NC QED.}
\end{figure}
\newpage
\begin{figure}[htb]
\centerline{ \DESepsf(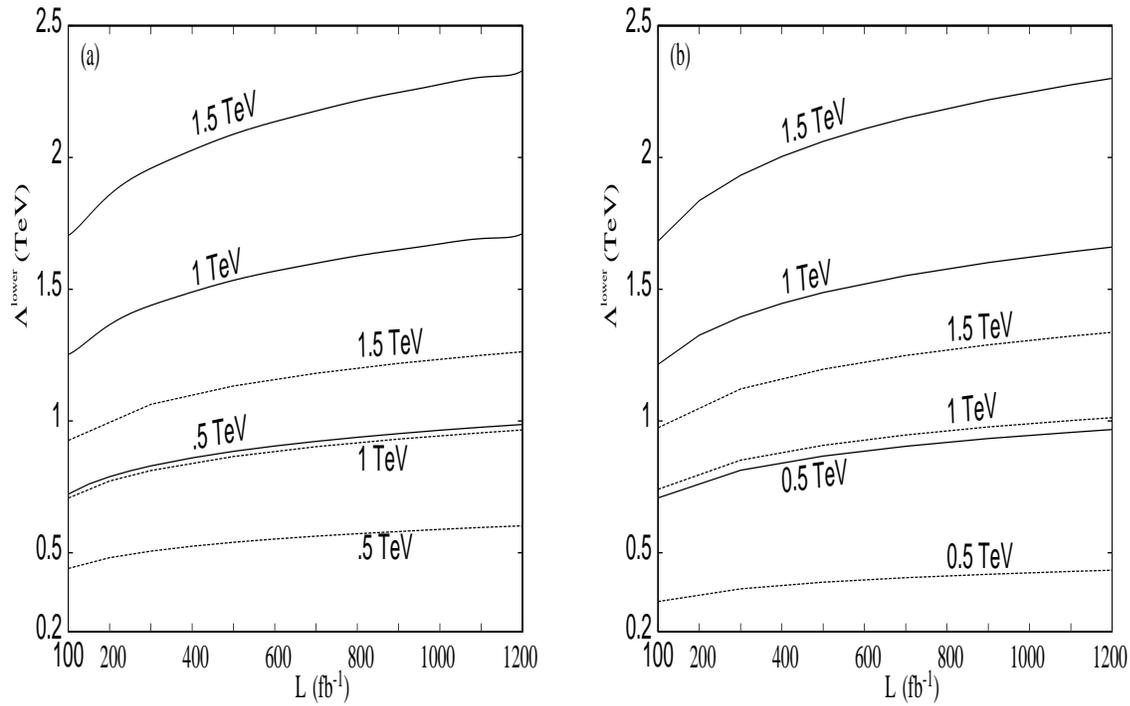 width 12cm)}
\caption {$95\%$ CL lower bound on $\Lambda $ can be probed 
as a function of integrated 
luminosity from $\gamma\gamma \rightarrow H^+ H^-$ process  for 
$m_H = 100$ GeV $(a)$ and $200$ GeV $(b)$. The solid lines are using 
monochromatic photon beams, while the dashed lines are with
back-scattered photons. The numbers adjacent to each curve represents the
$\sqrt{s}_{e^+e^-}$. }
\end{figure}

\newpage
\def\pr#1,#2,#3 { {\em Phys.~Rev.}        ~{\bf #1},  #2 (#3) }
\def\prd#1,#2,#3{ {\em Phys.~Rev.}        ~{\bf D#1}, #2 (#3) }
\def\prl#1,#2,#3{ {\em Phys.~Rev.~Lett.}  ~{\bf #1},  #2 (#3) }
\def\plb#1,#2,#3{ {\em Phys.~Lett.}       ~{\bf B#1}, #2 (#3) }
\def\npb#1,#2,#3{ {\em Nucl.~Phys.}       ~{\bf B#1}, #2 (#3) }
\def\prp#1,#2,#3{ {\em Phys.~Rept.}       ~{\bf #1},  #2 (#3) }
\def\zpc#1,#2,#3{ {\em Z.~Phys.}          ~{\bf C#1}, #2 (#3) }
\def\epj#1,#2,#3{ {\em Eur.~Phys.~J.}     ~{\bf C#1}, #2 (#3) }
\def\mpl#1,#2,#3{ {\em Mod.~Phys.~Lett.}  ~{\bf A#1}, #2 (#3) }
\def\ijmp#1,#2,#3{{\em Int.~J.~Mod.~Phys.}~{\bf A#1}, #2 (#3) }
\def\ptp#1,#2,#3{ {\em Prog.~Theor.~Phys.}~{\bf #1},  #2 (#3) }
\def\jhep#1,#2,#3{ {\em Journal of High Energy Physics}~{\bf #1}, #2 (#3) }


\begin{references}
\bibitem{snyder} 
H.~S.~Snyder, {\em Phys.~Rev.}~{\bf 71},38 (1947); 
{\em ibid},~{\bf 72},68 (1947);
\bibitem{nc_string} A.~Connes, {\it Non-commutative Geometry}, Academic Press,
(1994);
M.~Li and T.~Yoneya, \prl78,1219,{1997};
C.-S. Chu and P.-M. Ho, \npb550,151,{1999}.
A. Connes, M.R. Douglas and A. Schwarz, \jhep9802,003,{1998};
M.R. Douglas and C. Hull, \jhep9802,008,{1998};
N.~Seiberg and E.~Witten, \jhep9909,032,{1999};
M.M. Sheikh-Jabbari, \plb455,129,{1999} ;
J.L.F. Barbon and E. Rabinovici, \plb486,202,{2000};
R. Gopakumar, J. Maldacena, S. Minwalla and A. Strominger, \jhep0006,036,{2000};
D. Bigatti and L. Susskind, \prd62,066004,{2000};
N. Seiberg, L. Susskind and N. Toumbas, hep-th/0005040;
D.J. Gross, A. Hashimoto and N. Itzhaki, hep-th/0008075;
T. Pengpan and X. Xiong, hep-th/0009070;
F.~J.~Petriello,
hep-th/0101109.

\bibitem{extra} 
Nima Arkani-Hamed, Savas Dimopoulos and Gia Dvali, \plb429,263,{1998} ;
I.~Antoniadis, \plb246,377,{1990};
I.~Antoniadis, C.~Mu\~noz and M.~Quiros, \npb397, 515,{1993};
I.~Antoniadis, K.~Benakli and M.~Quiros, \plb331,313,{1994};
I.~Antoniadis, N.~Arkani-Hamed, S.~Dimopoulos and G.~Dvali, \plb463,257,{1998}.

\bibitem{arfaei}C.~P.~Martin, D.~Sanchez-Ruiz, \prl83,476,{1999};
 M.~Hayakawa, \plb478,394,{2000}, hep-th/9912094,{\em ibid}, 
hep-th/9912167; Ihab. F. Riad and M.M. Sheikh-Jabbari, \jhep0008,045,{2000};  
M. Chaichian, M.M. Sheikh-Jabbari and A. Tureanu, e-print hep-th/0010175;
H. Arfaei and M. H. Yavartanoo, e-print hep-th/0010244;

\bibitem{rizzo}J.~L.~Hewett, F.~J.~Petriello and T.~G.~Rizzo, hep-ph/0010354,\\
Prakash Mathews, hep-ph/0010354. 

\bibitem{jika} 
G.V. Jikia, and Yu.F. Pirogov, \plb,283,135,{1992}; 
L.-Z. Sun and Y.-Y. Lin, \prd,54,3563,{1996};
Surajit Chakrabarti, Debajyoti Choudhury, Rohini M. Godbole, and 
Biswarup Mukhopadhyaya, \plb434,347,{1998};
D. K. Ghosh, Prakash Mathews, P. Poulose, and K. Sridhar, 
\jhep9911,004,{1999};
Prakash Mathews, P. Poulose, and K. Sridhar, \plb461,196,{1999};
Xiao-Gang He, \prd60,11501,{2000};
Debajyoti Choudhury, and Anindya Datta, \npb592,35,{2001}.

\bibitem{matsubara} K. Matsubara, \plb482,417,{2000}.

\bibitem{nlc_rev} Hitoshi Murayama and Michael Peskin, 
Ann.~Rev.~Nucl.~Part.~Sci.~{\bf 46}, 533 (1996); 
E.~Accomando {\em et. al.} \prp299,1,{1998}.

\bibitem{riad} Ihab. F. Riad and M.M. Sheikh-Jabbari, \jhep0008,045,{2000}.

\bibitem{ginzburg} I. Ginzburg et al., 
Nucl. Instr. Methods {\bf 202}, 57(1983).

\bibitem{book} R. Field, Applications of Perturbative QCD, Addison-Wesley,
1989.

\end{references}
\end{document}